\documentclass
[floatfix,superscriptaddress,secnumarabic,amssymb,amsmath,nobibnotes,aps,prd,showkeys,showpacs,nofootinbib,onecolumn,notitlepage,10pt]{revtex4}%
\usepackage{setspace}
\usepackage{color}
\usepackage{amsmath}
\usepackage{amsfonts}
\usepackage{verbatim}
\usepackage{amssymb}
\usepackage{graphicx,bm}
\usepackage{graphicx}
\usepackage{amsmath}
\usepackage{amssymb}
\usepackage{amssymb}
\usepackage{graphicx,bm}
\usepackage{graphicx}
\usepackage[caption=false]{subfig}%
\providecommand{\U}[1]{\protect\rule{.1in}{.1in}}
\newcommand{\be}{\begin{equation}}
\newcommand{\ee}{\end{equation}}

\newcommand{\mincir}{\raise
-3.truept\hbox{\rlap{\hbox{$\sim$}}\raise4.truept\hbox{$<$}\ }}
\newcommand{\magcir}{\raise
-3.truept\hbox{\rlap{\hbox{$\sim$}}\raise4.truept\hbox{$>$}\ }}

\ifx\pdfoutput\relax\let\pdfoutput=\undefined\fi
\newcount\msipdfoutput
\ifx\pdfoutput\undefined\else
\ifcase\pdfoutput\else
\ifx\paperwidth\undefined\else
\ifdim\paperheight=0pt\relax\else\pdfpageheight\paperheight\fi
\ifdim\paperwidth=0pt\relax\else\pdfpagewidth\paperwidth\fi
\fi\fi\fi
\begin{document}
\title{Cosmological Bounds on the Generalized Uncertainty Principle from the CMB}
\author{Andronikos Paliathanasis}
\email{anpaliat@phys.uoa.gr}
\affiliation{Institute of Systems Science, Durban University of Technology, Durban 4000,
South Africa}
\affiliation{Centre for Space Research, North-West University, Potchefstroom 2520, South Africa}
\affiliation{Centro de Investigaci\'on, Innovaci\'on y Creaci\'on (CIIC), Universidad
Cat\'olica de Temuco, Temuco, Chile}
\affiliation{Departamento de Ciencias Matem\'{a}ticas y F\'{\i}sicas, Facultad de
Ingenier\'{\i}a, Universidad Cat\'olica de Temuco, Temuco, Chile}
\affiliation{National Institute for Theoretical and Computational Sciences (NITheCS), South Africa}

\begin{abstract}
We constrain the deformation parameter $\beta$ of the Generalized Uncertainty
Principle with cosmological observations. A quadratic deformation of the
Heisenberg uncertainty principle modifies the Friedmann equations of $\Lambda
$CDM and gives rise to a phenomenological dynamical dark energy with a single
free parameter, $\beta$. To avoid a singular behaviour for the effective dark
energy equation of state parameter, we restrict the prior to non-negative
values of $\beta$.
Using Planck 2018 CMB temperature and polarization data together with DESI DR2
BAO and Pantheon+SH0ES supernovae, we find the upper limit $\beta<0.00728~$ at
$68\%$ confidence interval. The resulting cosmological model fits the data
slightly better than $\Lambda$CDM, providing smaller value for the $\chi_{min}^{2}%
$, such that the Akaike Information Criterion to suggest statistical
equivalency between the two models, although the Bayesian evidence favours the
$\Lambda$CDM. Finally, the quadratic Generalized Uncertainty Principle does
not alleviate the $H_{0}$ tension, it reduces the tension to $4.8\sigma$, that
is, due to the small correlation between the $H_{0}-\beta$ parameters.

\end{abstract}
\keywords{Planck; Cosmological Constraints; Generalized Uncertainty Principle;
Deformation Parameter}\maketitle

\section{Introduction}

\label{sec1}

The existence of a maximum energy scale in nature or, equivalently, of a
minimal length of the order of the Planck length, $\ell_{\mathrm{Pl}}$, is
predicted by various approaches to quantum gravity, such as string theory
\cite{Veneziano:1986,Amati:1989} and doubly special relativity
\cite{Magueijo:2001cr}, as well as by gedanken experiments in black hole
physics \cite{Maggiore:1993rv,Scardigli:1999jh}; see also
\cite{Maggiore:1993zu,Garay:1994en,Ali:2009zq,Das:2010zf,Chung:2018btu}. The
existence of a minimal length is incompatible with the standard Heisenberg
Uncertainty Principle, which allows for arbitrarily small uncertainties in
position. A natural way to incorporate it into quantum mechanics is through a
deformation of the Heisenberg algebra, which leads to the Generalized
Uncertainty Principle (GUP)
\cite{kempH,Tawfik:2015rva,Scardigli:2014qka,Casadio:2020rsj,Valero:2025asd}.

The deformed algebra introduces correction terms in physical systems at
different scales \cite{kempH}, whose magnitude is determined by the
deformation parameter \cite{vag1}. The GUP corrections to the energy spectrum
of the hydrogen atom were derived in \cite{Brau:1999uv} and used to determine
an upper bound for the deformation parameter. The harmonic oscillator was
studied in \cite{Bosso:2017ndq}, while macroscopic mechanical oscillators have
been used to test the GUP experimentally \cite{Bawaj:2014cda,Bushev:2019zvw}.
An approach to constrain the deformation parameter with quantum optics was
proposed in \cite{Pikovski:2011zk}, while upper bounds were obtained from
gravitational bar detectors in \cite{Marin:2013pga}. Furthermore, tests of the
GUP with the gravitational-wave detector LIGO were proposed in
\cite{Bosso:2018ckz}. Astrophysical objects can also be used to test the GUP
\cite{Vagenas:2017vsw,Scardigli:2016pjs}, while observational data from the
Event Horizon Telescope have been used to bound the deformation parameter
\cite{Chen:2024mlr,Ali:2024ssf,Fragomeno:2026nnc}. For other applications, we
refer the reader to
\cite{Bera:2021lgw,Vagenas:2019rai,Heidari:2023ssx,Buoninfante:2020cqz,AlGhifari:2026vil,Naorem:2026cpr,Gulia:2025tvx,Gine:2025qsb,Nenmeli:2021orl}
and references therein.

In cosmological studies, the GUP introduces quantum corrections to the
Friedmann equations \cite{Vagenas:2017fwa}, which can account for the
acceleration of the cosmic expansion through an effective cosmological
constant term \cite{Vagenas:2019wzd}. Moreover, the deformation parameter
affects the inflationary epoch
\cite{Perivolaropoulos:2017rgq,Huang:2013byx,Tawfik:2012he,Heydarzadeh:2022lit,Tawfik:2014dza,Maziashvili:2011qs,Paliathanasis:2015cza,Heidarian:2025drk}%
, Big Bang nucleosynthesis \cite{Luciano:2021vkl,Luo:2024vdd}, as well as the
behavior of the universe as it approaches the cosmological singularity
\cite{Segreto:2024gcg}. The effects of the deformation parameter in
inhomogeneous cosmologies were examined in \cite{Paliathanasis:2025mfj}, where
it was shown that the deformed algebra modifies the Einstein field equations.
In the homogeneous limit, that is, for a
Friedmann-Lema\^{\i}tre-Robertson-Walker (FLRW) background, the deformed
algebra modifies the background dynamics of the $\Lambda$CDM model and gives
rise to a dynamical dark energy component, which drives the late-time cosmic
acceleration \cite{Paliathanasis:2025dcr}. Using the baryon acoustic
oscillation (BAO) measurements from the second data release of the Dark Energy
Spectroscopic Instrument (DESI DR2) \cite{DESI:2025zpo,DESI:2025zgx}, in
combination with Type Ia supernova (SNIa), cosmic chronometer, and
redshift-space distortion data, the deformation parameter was constrained for
the quadratic GUP in \cite{Paliathanasis:2025dcr,Paliathanasis:2026vhi}, and
for a more general power-law GUP framework in \cite{Paliathanasis:2025kmg}.

In this study, we extend the phenomenological analysis of the GUP-modified
$\Lambda$CDM model by employing the Cosmic Microwave Background (CMB) data
from the Planck 2018 release
\cite{Planck:2018nkj,Planck:2018vyg,Planck:2019nip} to constrain the
deformation parameter. Moreover, we investigate whether the GUP corrections
can alleviate the $H_{0}$ tension \cite{CosmoVerseNetwork:2025alb}, and hence
whether GUP-modified cosmological models deserve further investigation in the
context of the cosmological tensions. The paper is organized as follows.

In Section \ref{sec2}, we introduce the quadratic GUP and present the deformed
Friedmann equations for the $\Lambda$CDM model. We derive the closed-form
expression of the Hubble function for the GUP-modified theory and define the
effective equation of state parameter for the resulting dynamical dark energy
component. In Section \ref{sec3}, we present the observational constraints on
the model obtained from the CMB, BAO, and Pantheon+ SNIa data, and we examine
the deviation from the undeformed $\Lambda$CDM model. We also discuss the
constraints on the deformation parameter obtained from the CMB data. Finally,
in Section \ref{sec4}, we summarize our findings and draw our conclusions.

\section{Quadratic GUP}

\label{sec2}

We consider the quadratic GUP \cite{kempH,Vagenas:2020bys} where the position
$\Delta X$ and momentum $\Delta P$ uncertainties satisfy the inequality%
\begin{equation}
\Delta X\Delta P\geq\frac{1}{2}\left(  1+\beta\Delta P^{2}\right)  ,
\label{GUP}%
\end{equation}
and $\beta$ is the deformation parameter of dimension $1/P^{2}$. The latter
can be expressed in terms of the Planck mass,~$M_{Pl}$ and the Planck length
$\ell_{Pl}$ ($\approx10^{-35}~m$), as $\beta={\beta}_{0}/{M_{Pl}^{2}c^{2}=\beta
}_{0}\ell_{Pl}^{2}/2\hbar^{2}$, in which $M_{Pl}c^{2}$ ($\approx
1.2\times10^{19}~GeV)$ is the Planck energy and $\beta_{0}$ is a dimensionless
coupling parameter~\cite{Amati:1989}. The parameter $\beta$ is usually taken
to be positive; however, several approaches in the literature allow negative
values of the deformation parameter
\cite{Ong:2018zqn,Du:2022mvr,Buoninfante:2019fwr}.

The uncertainty conditions (\ref{GUP}) deform the relation for the canonical
algebra of the position and momentum as follows {\cite{Maggiore:1993rv}
\begin{equation}
\lbrack X_{i},P_{j}]=i\left[  \delta_{ij}(1+\beta P^{2})+2\beta P_{i}%
P_{j}\right]  , \label{gup2}%
\end{equation}
with }%
\begin{equation}
{\left[  X_{i},X_{j}\right]  =0\ }\text{{and}}{\ \left[  P_{i,}P_{j}\right]
=0.}%
\end{equation}

In terms of the standard operators $x_{i},~p_{i}$, with {$[x_{i},p_{j}%
]=i\hbar\delta_{ij}$\thinspace$,$ and $p^{2}=p_{i}p^{i},$ it follows%
\begin{equation}
P_{i}=(1+\beta p^{2})p_{i},
\end{equation}
The Poisson brackets are modified accordingly, and Hamilton's
equations for an autonomous system }$\mathcal{H}=\mathcal{H}(x,p)$ become
\cite{Bosso:2018uus,Bosso:2023aht}%
\begin{equation}
\dot{x}_{i}=\left(  1-\beta p^{2}\right)  \frac{\partial\mathcal{H}}{\partial
p_{i}},~\dot{p}_{i}=\left(  1-\beta p^{2}\right)  \frac{\partial\mathcal{H}%
}{\partial x_{i}}.
\end{equation}

\subsection{GUP-modified $\Lambda$CDM}

For the $\Lambda$CDM model, the minimal length modifies the second Friedmann
equation as
\[
\dot{H}+H^{2}=\frac{1}{6}\left(  2\Lambda-\rho_{m}\right)  \left(  1+2\beta
f\left(  t\right)  \right)  +\beta\dot{f}\left(  t\right)  H+O\left(
\beta^{2}\right)
\]
with deformation function $f\left(  t\right)  =\frac{H^{2}}{\rho_{m}}$, or
$f\left(  t\right)  =\frac{1}{3\Omega_{m}}$, with $\Omega_{m}=\frac{\rho_{m}%
}{3H^{2}}$. For more details, we refer the reader to
\cite{Paliathanasis:2025mfj}.

To first order in $\beta$, the matter density parameter $\Omega_{m}$
evolves as%
\[
\frac{d\Omega_{m}}{d\ln a}=3\Omega_{m}\left(  \Omega_{m}-1\right)  +\left(
4\Omega_{m}-\frac{10}{3}\right)  \beta,
\]
with solution \cite{Paliathanasis:2025dcr}
\begin{equation}
\Omega_{m}\left(  a\right)  =\frac{1}{6}\left(  3-4\beta+\sqrt{9+16\beta
}-\frac{2\sqrt{9+16\beta\left(  1+\beta\right)  }}{1+\left(  c_{1}\right)
^{-1}a^{-\sqrt{9+16\beta\left(  1+\beta\right)  }}}\right)  ,\label{fe.28}%
\end{equation}
from which the background Hubble function follows \cite{Paliathanasis:2025dcr}
\begin{equation}
H^{2}\left(  a\right)  =H_{G0}^{2}\frac{3\Omega_{m0}a^{-3}}{\left(
\frac{\sqrt{1+\frac{16}{9}\beta\left(  1+\beta\right)  }\Omega_{m0}a^{-3}%
}{\Omega_{m0}a^{-3}+\left(  1-\Omega_{m0}\right)  a^{3\left(  -1+\sqrt
{1+\frac{16}{9}\beta\left(  1+\beta\right)  }\right)  }}+\frac{1}{2}\left(
1-\sqrt{1+\frac{16}{9}\beta\left(  1+\beta\right)  }\right)  -\frac{2}{3}%
\beta\right)  }.\label{fe.29a}%
\end{equation}
We now write the Hubble function in the form
\begin{equation}
H^{2}\left(  a\right)  =H_{0}^{2}\left(  \Omega_{m0}a^{-3}+\Omega_{DE}\left(
a\right)  \right)  ,\label{fe.33aa}%
\end{equation}
with $\Omega_{DE}\left(  a\right)  =\Omega_{DE0}e^{-3\int_{a}^{a_{0}}%
\frac{\left(  1+w_{DE}\left(  a\right)  \right)  }{a}da}$. Hence,%
\begin{equation}
\Omega_{DE}\left(  a\right)  =\left(  \frac{H\left(  a\right)  }{H_{0}%
}\right)  ^{2}-\Omega_{m0}a^{-3},\label{fe1}%
\end{equation}
and
\begin{equation}
w_{DE}\left(  a\right)  =-1-\frac{1}{3}\frac{d}{d\ln a}\ln\left(  \left(
\frac{H\left(  a\right)  }{H_{0}}\right)  ^{2}-\Omega_{m0}a^{-3}\right)
.\label{fe2}%
\end{equation}
We remark that, in the absence of the deformation parameter, that is, $\beta\rightarrow0$, we obtain the $\Lambda$CDM limit.

In the following, we use this GUP-modified dark energy equation of state
parameter to derive cosmological constraints and to determine bounds on the
deformation parameter $\beta$. Furthermore, we assume that the quantum
corrections affect the perturbation equations only indirectly, through the
modified background Hubble function. Direct GUP corrections to the
perturbation equations are of order $O(\beta^{2})$ or $O(\beta\delta\rho)$,
and they can be neglected. Therefore, the perturbations evolve according to
General Relativity on the GUP-modified background\footnote{A limitation of the present analysis is that the closed-form expression for
$w_{DE}(z)$ was derived in the absence of radiation. Although radiation is
expected to modify the evolution of $w_{DE}(z)$ at early times, in this work
we assume that expression (\ref{fe2}) remains valid throughout the entire
cosmic history. This assumption is a technical simplification, which allows us
to implement the GUP-modified dark energy component directly in the
\texttt{CLASS} code \cite{Blas:2011rf}.}.

\section{CMB Constraints on the quadratic GUP}

\label{sec3}

To constrain the dynamical dark-energy model (\ref{fe2}) with the
observational data, we employ \texttt{Cobaya}%
\footnote{https://cobaya.readthedocs.io/} \cite{cob1,cob2} with the Boltzmann
solver \texttt{CLASS} \cite{Blas:2011rf} and the MCMC sampler
\cite{mcmc1,mcmc2}. Moreover, for the analysis of the numerical chains we
employ the \texttt{GetDist} package \cite{getd}. 

We consider the following data sets

\begin{itemize}
\item SNIa: We consider the Pantheon+ (PP) SNIa catalogue
\cite{Brout:2022vxf}, which contains 1550 SNIa observations in the redshift
range $10^{-3}<z<2.27$. In addition, for the background constraints we
consider the PP sample calibrated with the SH0ES Cepheids, which allow direct
measurement of the $H_{0}$.

\item BAO: We make use of the latest BAO measurements from DESI DR2
\cite{DESI:2025zgx,DESI:2025zpo}, which are given in seven redshift bins at
the range $0.29<z<2.34$, in terms of the transverse comoving distance
$D_{M}r_{drag}^{-1},~$the volume-averaged distance ratio, $D_{V}r_{drag}^{-1}$
and the Hubble distance ratio $D_{H}r_{drag}^{-1}$, where $r_{drag}$ is
the sound horizon at the baryon drag epoch.

\item CMB: Finally, we employ the Planck 2018 low-$\ell$ temperature and
E-mode polarization likelihoods \cite{Planck:2019nip}, and the high-$\ell$
TT,TE,EE likelihood for $\ell\geq30$, which we refer to as Planck 2018 TT,TE,EE+lowE.
\end{itemize}

For this investigation, we consider the following data combinations, (i)
CMB+BAO, (ii) CMB+BAO+PP, and, using only background data, (iii) BAO+PP+SH0ES.
The latter combination allows us to examine whether the GUP-modified model can
alleviate the $H_{0}$ tension. The priors of the cosmological parameters
considered in the numerical analysis are presented in Table~\ref{priors}. For
the deformation parameter, we consider only positive values, since for
$\beta<0$ the dark energy density $\Omega_{DE}(a)$ changes sign during the
cosmic evolution and the equation of state parameter $w_{DE}(a)$ becomes
singular for large redshifts, see the discussion in
\cite{Paliathanasis:2025dcr}. For the background data combinations (iii),
the sound horizon at the baryon drag epoch, $r_{drag}$ is treated as a free
parameter with prior $r_{drag}\in\left[  120,160\right]  $.%

%TCIMACRO{\TeXButton{B}{\begin{table}[tbp] \centering}}%
%BeginExpansion
\begin{table}[tbp] \centering
%EndExpansion
\caption{Priors of the cosmological parameters.}%
\begin{tabular}
[c]{cc}\hline\hline
\textbf{Parameters} & \textbf{Prior}\\\hline
$\mathbf{H}_{0}$ & $[60,\,80]$\\
$\mathbf{\Omega}_{b}\mathbf{h}^{2}$ & $[0.005,\,0.1]$\\
$\mathbf{\Omega}_{dm}\mathbf{h}^{2}$ & $[0.001,\,0.99]$\\
$\mathbf{\tau}$ & $[0.01,\,0.8]$\\
$\mathbf{n}_{s}$ & $[0.8,\,1.2]$\\
$\ln\left(  10^{10}A_{s}\right)  $ & $[1.61,\,3.91]$\\
$\mathbf{\beta}$ & $[0,\,0.1]$\\\hline\hline
\end{tabular}
\label{priors}%
%TCIMACRO{\TeXButton{E}{\end{table}}}%
%BeginExpansion
\end{table}%
%EndExpansion

\subsection{Results}

\label{sec4a}

We continue with the presentation of the observational constraints. In Tables
\ref{tabl1} and \ref{tabl2} we present the marginalized posterior medians for
the cosmological parameters with their $68\%$ and $95\%~$credible intervals
(CI) for the modified GUP theory and the $\Lambda$CDM, which is used as
reference model. Table \ref{tabl1} refers to the CMB+BAO and CMB+BAO+PP combinations, and
Table \ref{tabl2} refers to the BAO+PP+SH0ES dataset. Furthermore, in Table
\ref{tabten} we show the tension for the obtained cosmological parameters
between the two models. To compare the statistical significance of the two
models, we employ the Akaike Information Criterion \cite{AIC} and the Bayesian
evidence\footnote{For the derivation of the Bayesian evidence from the MCMC chains we employ the \texttt{MCEvidence} package \cite{Heavens:2017afc}} \cite{AIC2}.

%TCIMACRO{\TeXButton{B}{\begin{table}[tbp] \centering}}%
%BeginExpansion
\begin{table}[tbp] \centering
%EndExpansion
\caption{Marginalized constraints on the model parameters, together with their posterior mean at 68\% and 95\% credible intervals, for the modified GUP and $\Lambda$CDM models obtained from different combinations of the observational data (1/2).}%
\begin{tabular}
[c]{ccccc}\hline\hline
\textbf{Parameter} & \multicolumn{2}{c}{\textbf{CMB+BAO}} &
\multicolumn{2}{c}{\textbf{CMB+BAO+PP}}\\\hline
& \textbf{GUP} & $\Lambda$\textbf{CDM} & \textbf{GUP} & $\Lambda$%
\textbf{CDM}\\\hline
$\mathbf{H}_{0}$ & $68.79_{-0.51\left(  -0.87\right)  }^{+0.30\left(
+0.92\right)  }$ & $68.45_{-0.31\left(  -0.61\right)  }^{+0.31\left(
+0.60\right)  }$ & $68.69_{-0.40\left(  -0.73\right)  }^{+0.33\left(
+0.77\right)  }$ & $68.45_{-0.30\left(  -0.58\right)  }^{+0.30\left(
+0.59\right)  }$\\
$\mathbf{\Omega}_{m0}$ & $0.298_{-0.0040\left(  -0.0094\right)  }%
^{+0.0051\left(  +0.0095\right)  }$ & $0.300_{-0.004\left(  -0.0075\right)
}^{+0.004\left(  +0.0078\right)  }$ & $0.298_{-0.0043\left(  -0.0083\right)
}^{+0.0043\left(  +0.0083\right)  }$ & $0.3003_{-0.0038\left(  -0.0074\right)
}^{+0.0038\left(  +0.0076\right)  }$\\
$\mathbf{r}_{drag}$ & $147.54_{-0.23\left(  -0.42\right)  }^{+0.23\left(
+0.42\right)  }$ & $147.60_{-0.20\left(  -0.41\right)  }^{+0.20\left(
+0.42\right)  }$ & $147.54_{-0.22\left(  -0.41\right)  }^{+0.22\left(
+0.41\right)  }$ & $147.60_{-0.20\left(  -0.40\right)  }^{+0.20\left(
+0.39\right)  }$\\
$\mathbf{\sigma}_{8}$ & $0.8041_{-0.0076\left(  -0.015\right)  }%
^{+0.0065\left(  +0.015\right)  }$ & $0.8041_{-0.0071\left(  -0.014\right)
}^{+0.0071\left(  +0.014\right)  }$ & $0.8040_{-0.0074\left(  -0.014\right)
}^{+0.0066\left(  +0.015\right)  }$ & $0.8042_{-0.0071\left(  -0.014\right)
}^{+0.0071(+0.014)}$\\
$\mathbf{S}_{8}$ & $0.801_{-0.011\left(  -0.021\right)  }^{+0.011\left(
+0.021\right)  }$ & $0.8045_{-0.0098\left(  -0.019\right)  }^{+0.0098\left(
+0.020\right)  }$ & $0.802_{-0.010\left(  -0.020\right)  }^{+0.010\left(
+0.020\right)  }$ & $0.8046_{-0.0097\left(  -0.019\right)  }^{+0.0097\left(
+0.019\right)  }$\\\hline
$\mathbf{\beta}$ & $<0.00728\left(  0.0190\right)  $ & $-$ & $<0.00618\left(
0.0146\right)  $ & $-$\\\hline
$\mathbf{\Omega}_{b}\mathbf{h}^{2}$ & $0.02252_{-0.0.00014\left(
-0.00025\right)  }^{+0.0.00014\left(  +0.00025\right)  }$ &
$0.02252_{-0.00013\left(  -0.00025\right)  }^{+0.00013\left(  +0.00025\right)
}$ & $0.02252_{-0.00013\left(  -0.00025\right)  }^{+0.00013\left(
+0.00025\right)  }$ & $0.02252_{-0.00013\left(  -0.00025\right)
}^{+0.00013\left(  +0.00025\right)  }$\\
$\mathbf{\Omega}_{c}\mathbf{h}^{2}$ & $0.11759_{-0.00071\left(
-0.0014\right)  }^{+0.00071\left(  +0.0014\right)  }$ &
$0.11751_{-0.00068\left(  -0.0013\right)  }^{+0.00068\left(  +0.0014\right)
}$ & $0.11760_{-0.00069\left(  -0.0014\right)  }^{+0.00069\left(
+0.0014\right)  }$ & $0.11751_{-0.00067\left(  -0.0013\right)  }%
^{+0.00067\left(  +0.0013\right)  }$\\
$\mathbf{n}_{s}$ & $0.9707_{-0.0034\left(  -0.0067\right)  }^{+0.0034\left(
+0.0067\right)  }$ & $0.9706_{-0.0034\left(  -0.0066\right)  }^{+0.0034\left(
+0.0066\right)  }$ & $0.9706_{-0.0034\left(  -0.0066\right)  }^{+0.0034\left(
+0.0066\right)  }$ & $0.9706_{-0.0034\left(  -0.0066\right)  }^{+0.0034\left(
+0.0066\right)  }$\\
$\ln\left(  10^{10}A_{s}\right)  $ & $3.046_{-0.017\left(  -0.034\right)
}^{+0.017\left(  +0.034\right)  }$ & $3.045_{-0.016\left(  -0.032\right)
}^{+0.016\left(  +0.034\right)  }$ & $3.046_{-0.017\left(  -0.032\right)
}^{+0.016\left(  +0.035\right)  }$ & $3.046_{-0.017\left(  -0.034\right)
}^{+0.017\left(  +0.034\right)  }$\\
$\mathbf{\tau}$ & $0.0581_{-0.0083\left(  -0.015\right)  }^{+0.0073\left(
+0.017\right)  }$ & $0.0576_{-0.0079\left(  -0.015\right)  }^{+0.0079\left(
+0.016\right)  }$ & $0.0580_{-0.0083\left(  -0.015\right)  }^{+0.0073\left(
+0.017\right)  }$ & $0.0577_{-0.0082\left(  -0.015\right)  }^{+0.0073\left(
+0.016\right)  }$\\\hline
$\mathbf{\Delta\chi}_{\min}^{2}$ & $-1.36$ & $0$ & $-0.34$ & $0$\\
$\mathbf{\Delta AIC}$ & $+0.64$ & $0$ & $+1.66$ & $0$\\
$\mathbf{\Delta}\ln\mathbf{Z}$ & $-1.78$ & $0$ & $-0.84$ & $0$\\\hline\hline
\end{tabular}
\label{tabl1}%
%TCIMACRO{\TeXButton{E}{\end{table}}}%
%BeginExpansion
\end{table}%
%EndExpansion
%

%TCIMACRO{\TeXButton{B}{\begin{table}[tbp] \centering}}%
%BeginExpansion
\begin{table}[tbp] \centering
%EndExpansion
\caption{Marginalized constraints on the model parameters, together with their posterior means at 68\% and 95\% credible intervals, for the modified GUP and $\Lambda$CDM models obtained from different combinations of the observational data (2/2).}%
\begin{tabular}
[c]{ccc}\hline\hline
\textbf{Parameter} & \multicolumn{2}{c}{\textbf{BAO+PP+SH0ES}}\\\hline
& \textbf{GUP} & $\Lambda$\textbf{CDM}\\\hline
$\mathbf{H}_{0}$ & $73.8_{-1.0\left(  -2.1\right)  }^{+1.0\left(  +2.0\right)
}$ & $73.8_{-1.0\left(  -1.9\right)  }^{+1.0\left(  +2.0\right)  }$\\
$\mathbf{\Omega}_{m0}$ & $0.319_{-0.014\left(  -0.028\right)  }^{+0.014\left(
+0.030\right)  }$ & $0.304_{-0.012\left(  -0.023\right)  }^{+0.012\left(
+0.024\right)  }$\\
$\mathbf{r}_{drag}$ & $136.7_{-2.3\left(  -4.5\right)  }^{+2.3\left(
+4.3\right)  }$ & $136.9_{-2.3\left(  -4.3\right)  }^{+2.3\left(  +4.7\right)
}$\\\hline
$\mathbf{\beta}$ & $<0.0104\left(  0.0232\right)  $ & $-$\\\hline
$\mathbf{\Delta\chi}_{\min}^{2}$ & $+0.00$ & $0$\\
$\mathbf{\Delta AIC}$ & $+2.00$ & $0$\\
$\mathbf{\Delta}\ln\mathbf{Z}$ & $-2.84$ & $0$\\\hline\hline
\end{tabular}
\label{tabl2}%
%TCIMACRO{\TeXButton{E}{\end{table}}}%
%BeginExpansion
\end{table}%
%EndExpansion

\begin{figure}[ptbh]
\centering\includegraphics[width=1\textwidth]{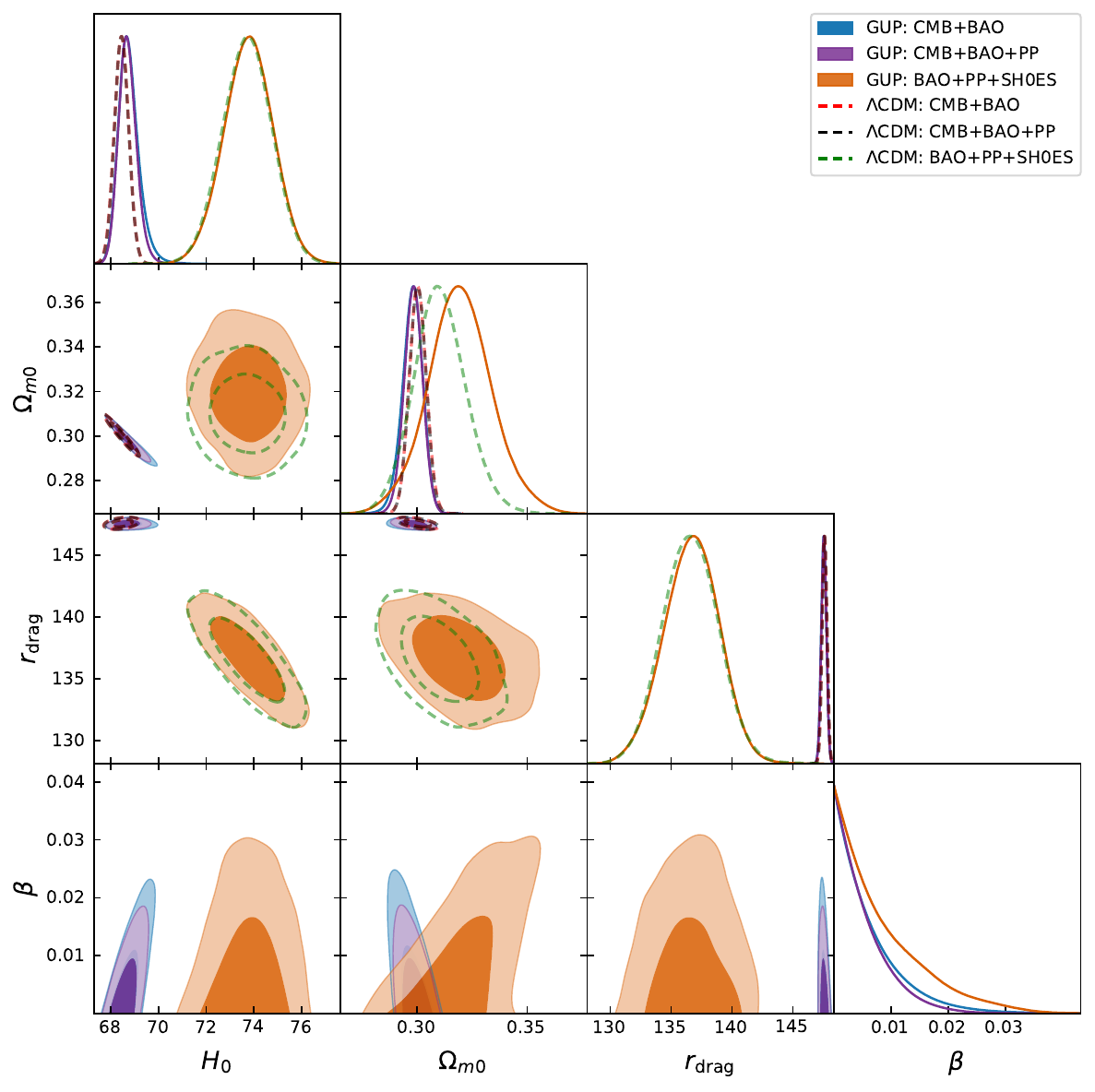}\caption{Marginalized
distributions for the background cosmological parameters $H_{0}$, $\Omega
_{m0}$, $r_{drag}$ and $\beta$. The contours correspond to the $68\%$ and
$95\%$ credible regions. Filled contours correspond to the modified GUP model,
and dashed lines to the $\Lambda$CDM. }%
\label{fig1}%
\end{figure}

For the CMB+BAO combination, the Hubble constant is $H_{0}=68.79_{-0.5}%
^{+0.30}~\frac{km~s^{-1}}{Mpc}$ slightly larger than the $\Lambda$CDM value
$H_{0}=68.45_{-0.31}^{+0.31}\frac{km~s^{-1}}{Mpc}$, but consistent with it within
$0.6\sigma$. The same holds when the PP SNIa catalogue is introduced where
the obtained values are different only by $0.5\sigma$. By comparing the
CMB+BAO constraints with those from the BAO+PP+SH0ES combination, the $H_{0}$
tension is $4.8\sigma$, only $0.3\sigma$ below the $5.1\sigma$ obtained for
$\Lambda$CDM. Therefore, the quadratic GUP model does not alleviate the Hubble
tension; it only shifts $H_{0}$ marginally upward. In Fig. \ref{fig1} we
show the marginalized distributions for the background cosmological parameters
$H_{0},~\Omega_{m0},~r_{drag}$ and $\beta$ for the GUP modified model and the
$\Lambda$CDM for the three different combined datasets. We observe that the
larger uncertainty in $H_{0}$ is due to its weak correlation with the
deformation parameter $\beta$.

Furthermore, for the $\Omega_{m0}$ we find that the modified GUP model
provides a larger value for the background in comparison to the $\Lambda$CDM,
but within the $0.8\sigma$. For the background data, difference is less than
$0.4\sigma$, with the quadratic GUP model to admit smaller systematically
smaller values. Finally, for the rest of the cosmological parameters the
obtained values is statistically equivalent with difference less than
$0.2\sigma$.

Moreover, for the matter energy density $\Omega_{m0}$, the modified GUP model
gives larger value than $\Lambda$CDM for the BAO+PP+SH0ES combination,
although the difference is only $0.8\sigma$. For the CMB-based combinations
the difference is at most $0.4\sigma$, with GUP giving a slightly smaller
value. Finally, the remaining cosmological parameters are statistically
equivalent in the two models, with differences of at most $0.2\sigma$

For the deformation parameter $\beta$ we find that it is constrained only from
above. For the background data BAO+PP+SH0ES we find $\beta<0.0104$ within the
$68\%$ CI. However, when the CMB data are considered we find that
$\beta<0.00728$ for the CMB+BAO dataset, and $\beta<0.00618$ for the
CMB+BAO+PP data combination.

Finally, as far as the comparison of the statistical parameters for the two
models is concerned, when CMB data are included, the quadratic GUP model
provides a slightly better fit with a smaller $\chi_{\min}^{2}$. Indeed, for the CMB+BAO data
$\Delta\chi_{\min}^{2}=-1.36$, and for the CMB+BAO+PP combination $\Delta
\chi^{2}=-0.34$, with corresponding $\Delta AIC$ values of $+0.64$ and
$+1.66$. Hence, according to the Akaike scale \cite{AIC}, the models are
statistically equivalent. The same conclusion holds for the background
data BAO+PP+SH0ES. For the background data the difference of the Bayesian
evidence ${\Delta}\ln\mathbf{Z}$ reads $-2.84$, indicating a preference for
$\Lambda$CDM on the Jeffreys scale \cite{AIC2}. For the CMB+BAO data
set, we find $\Delta\ln Z=-1.78$, which leads to the same conclusion as before.
Nevertheless, for the CMB+BAO+PP combination we find $\Delta\ln Z=-0.84$,
i.e., the two models are statistically indistinguishable.

\begin{figure}[ptb]
\centering
\includegraphics[width=1\textwidth]{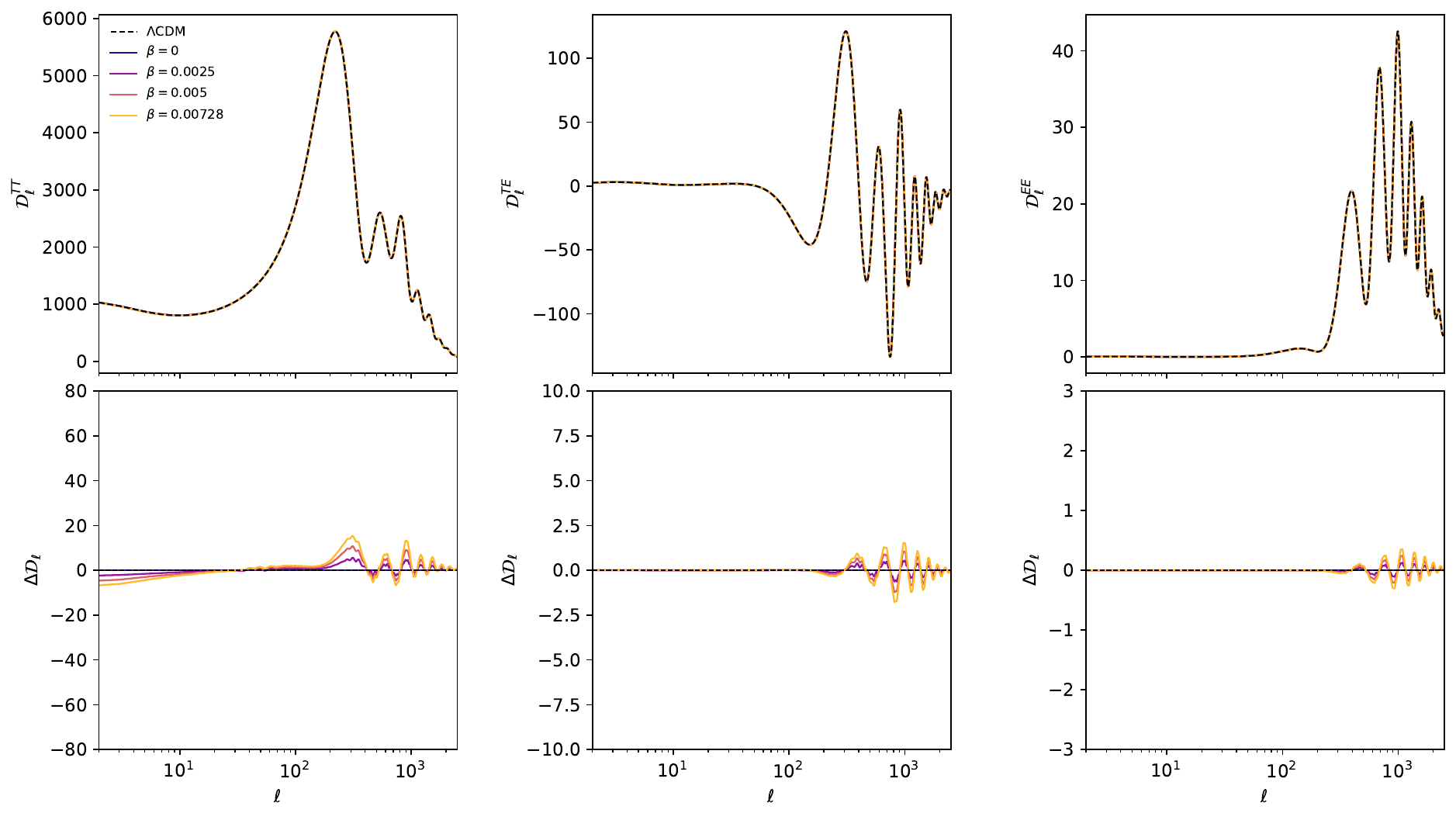} \caption{TT $\left(  D_{\ell
}^{TT}\right)  $, TE~$\left(  D_{\ell}^{TE}\right)  $, and EE $\left(
D_{\ell}^{EE}\right)  $ band powers for the quadratic GUP model for different
values of the deformation parameter $\beta$. The dashed black line is
$\Lambda$CDM $\left(  \beta=0\right)  $. The difference $\Delta\mathcal{D}%
_{\ell}=\mathcal{D}_{\ell}-\mathcal{D}_{\ell}^{\Lambda\mathrm{CDM}}$ between
the $\beta\neq0$ and $\beta=0$ models is also shown. Non-zero differences
appear at high $\ell$. }%
\label{fig2}%
\end{figure}

\begin{figure}[ptb]
\centering
\includegraphics[width=0.8\columnwidth]{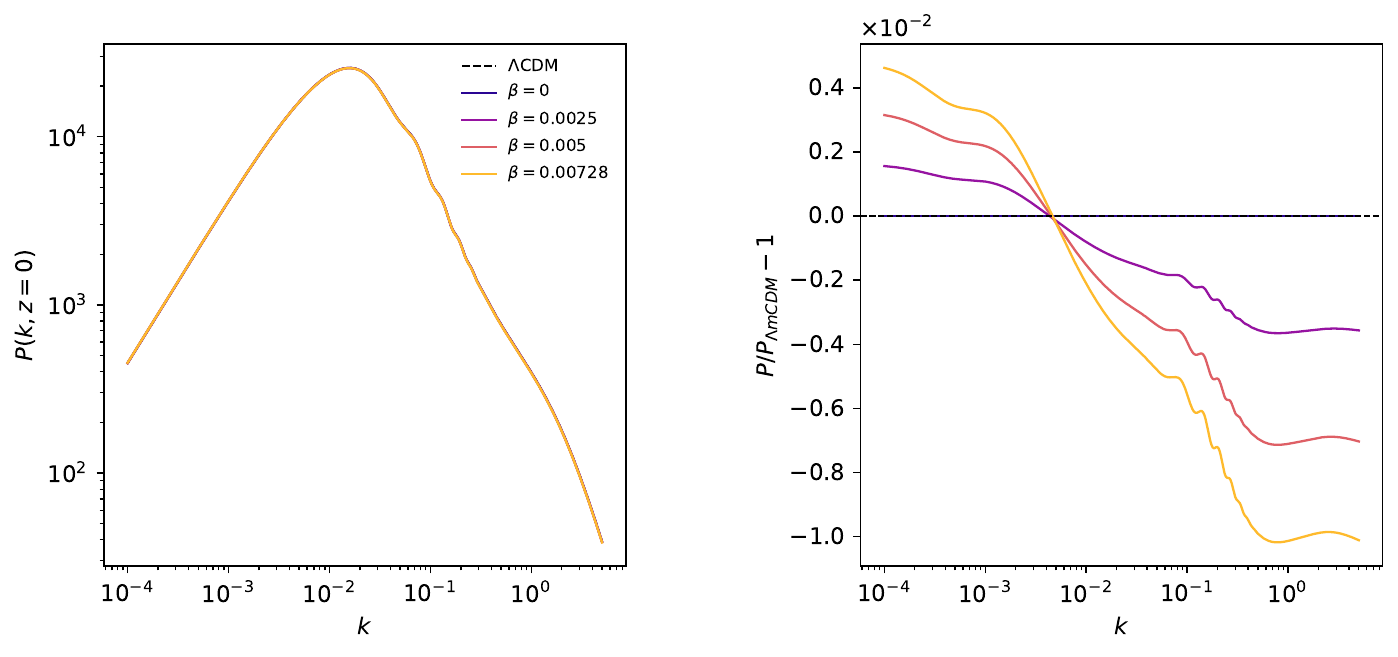} \caption{Matter spectra
$\left(  P\left(  k\right)  \right)  ~$at the present for the quadratic GUP
model for different values of $\beta$ and comparison to the $\Lambda$CDM. }%
\label{fig3}%
\end{figure}

In Figs. \ref{fig2} and \ref{fig3} we present the TT $\left(  D_{\ell}%
^{TT}\right)  $, TE~$\left(  D_{\ell}^{TE}\right)  $, and EE $\left(  D_{\ell
}^{EE}\right)  $ band powers and the Power spectra $P\left(  k\right)  $ at
the present, for the quadratic GUP model for different value of the
deformation parameter $\beta$, and we compare it with the $\Lambda$CDM limit.
The effects of the deformation parameter are small compared with the
predictions of $\Lambda$CDM resulting in negligible changes to the posterior
distributions of the corresponding cosmological parameters, as summarized in
Table \ref{tabten}.%

%TCIMACRO{\TeXButton{B}{\begin{table}[tbp] \centering}}%
%BeginExpansion
\begin{table}[tbp] \centering
%EndExpansion
\caption{Tension of the cosmological parameters for the modified GUP and $\Lambda$CDM models for each dataset.}
\begin{tabular}
[c]{lccc}\hline\hline
\textbf{Parameter} & \textbf{CMB+BAO} & \textbf{CMB+BAO+PP} &
\textbf{BAO+PP+SH0ES}\\\hline
\multicolumn{1}{c}{$\mathbf{H}_{0}$} & $0.6\sigma$ & $0.5\sigma$ & $0.0\sigma
$\\
\multicolumn{1}{c}{$\mathbf{\Omega}_{m0}$} & $0.3\sigma$ & $0.4\sigma$ &
$0.8\sigma$\\
\multicolumn{1}{c}{$\mathbf{r}_{\mathrm{drag}}$} & $0.2\sigma$ & $0.2\sigma$ &
$0.1\sigma$\\
\multicolumn{1}{c}{$\mathbf{\sigma}_{8}$} & $0.0\sigma$ & $0.0\sigma$ & --\\
\multicolumn{1}{c}{$\mathbf{S}_{8}$} & $0.2\sigma$ & $0.2\sigma$ & --\\
\multicolumn{1}{c}{$\mathbf{\Omega}_{b}\mathbf{h}^{2}$} & $0.0\sigma$ &
$0.0\sigma$ & --\\
\multicolumn{1}{c}{$\mathbf{\Omega}_{c}\mathbf{h}^{2}$} & $0.1\sigma$ &
$0.1\sigma$ & --\\
\multicolumn{1}{c}{$\mathbf{n}_{s}$} & $0.0\sigma$ & $0.0\sigma$ & --\\
\multicolumn{1}{c}{$\ln\mathbf{(10}^{10}\mathbf{A}_{s}\mathbf{)}$} &
$0.0\sigma$ & $0.0\sigma$ & --\\
\multicolumn{1}{c}{$\mathbf{\tau}$} & $0.0\sigma$ & $0.0\sigma$ &
--\\\hline\hline
\end{tabular}
\label{tabten}
%TCIMACRO{\TeXButton{E}{\end{table}}}%
%BeginExpansion
\end{table}%
%EndExpansion

\section{Conclusions}

\label{sec4}

In this study, we constrained the deformation parameter $\beta$ of the
quadratic GUP using the Planck 2018 CMB measurements. The deformed algebra
associated with the quadratic GUP modifies the Friedmann equations of the
$\Lambda$CDM model, such that the resulting theory is equivalent to a
dynamical dark energy model with a single free parameter, the deformation
parameter $\beta$. Moreover, assuming that the deformed algebra does not
affect the cosmological perturbation equations at first order, we implemented
this dynamical dark energy model as a $w\left(  a\right)  $CDM model in
the Boltzmann solver \texttt{CLASS}, and we used \texttt{Cobaya} to derive
constraints for different data combinations.

The standard $\Lambda$CDM theory is recovered from the quadratic GUP model
when $\beta=0$. Therefore, we adopted priors for the cosmological parameters
similar to those of $\Lambda$CDM. The deformation parameter was assigned the
positive prior $\beta\geq0$, because for negative values, $\beta<0$, the
effective energy density of the dark energy can become negative and the
effective equation of state parameter becomes singular.

For the background data BAO+PP+SH0ES, the deformation parameter is bounded
from above as $\beta<0.0104\,(0.0232)$ at the $68\%\,(95\%)$ CI. However, when
the Planck CMB data are included, we find $\beta<0.00728\,(0.0190)$ for the
CMB+BAO combination and $\beta<0.00618\,(0.0146)$ when the PP SNIa catalogue,
without the SH0ES calibration, is added. The remaining cosmological parameters
are practically unchanged from their $\Lambda$CDM values. Nevertheless, owing
to the weak correlation between the deformation parameter and the Hubble
constant $H_{0}$, the latter is shifted to a slightly higher value for the CMB
data, $H_{0}=68.79 ~\frac{km~s^{-1}}{Mpc}$, which lies within $0.6\sigma$ of
the corresponding $\Lambda$CDM value. This reduces the $H_{0}$ tension with
the background data by only $0.3\sigma$, from which we conclude that the
quadratic GUP does not alleviate the $H_{0}$ tension.

In our analysis, the equation of state of the phenomenological dynamical dark
energy was derived for the late universe, where radiation can be neglected,
and we assumed that this expression holds throughout the cosmic history. A
complete treatment would include radiation in the deformed Friedmann equations
and investigate the effects of the deformation parameter during the radiation
era. Although the quadratic GUP cannot resolve the cosmological tensions and
the cosmological data do not show a preference for a nonzero $\beta$, this
study opens a new direction: investigating whether higher-order or
non-quadratic GUP deformations are supported by cosmological observations and
can address the cosmological tensions.

\begin{acknowledgments}
Part of this study was supported by FONDECYT 1240514. The numerical analysis
was carried out using the computing cluster of the Universidad Cat\'olica de Temuco.
AP acknowledges the support of the COST Action CA23130 \textquotedblleft Bridging high and
low energies in search of quantum gravity (BridgeQG)\textquotedblright.
\end{acknowledgments}

\bibliography{biblio}

\end{document}